\documentclass{article}

\usepackage{preprint}
\usepackage{upgreek}				
\usepackage[utf8]{inputenc} 		
\usepackage[T1]{fontenc}    		
\usepackage[dvipsnames]{xcolor}	
\usepackage{hyperref}       		
\hypersetup{
    colorlinks = true,
    linkcolor = {NavyBlue},		
    anchorcolor = .,
    citecolor = {NavyBlue},		
    filecolor = .,				
    menucolor = .,
    runcolor = {cyan}, 
    urlcolor = {NavyBlue},		
    }
\usepackage{url}            
\usepackage{booktabs}       
\usepackage{amsfonts}       
\usepackage{nicefrac}       
\usepackage{microtype}      
\usepackage{fancyhdr}       
\usepackage{graphicx}       
\usepackage{setspace} 
\usepackage{soul} 
\usepackage{ragged2e} 
\usepackage[toc,page]{appendix}
\usepackage{longtable}
\usepackage{eso-pic}
\usepackage{tikz}
\usepackage[superscript]{cite}

\graphicspath{{./Figures/}}     

  \centering
\title{Hardware Implementation of Tunable Fractional-Order Capacitors by Morphogenesis of Conducting Polymer Dendrites}

\author{
  Antoine Baron\textsuperscript{a}, 
  Enrique H. Balaguera\textsuperscript{b}, 
  Corentin Scholaert\textsuperscript{a}, 
  Fabien Alibart\textsuperscript{c}, 
  \\
  \bf{S\'{e}bastien Pecqueur\textsuperscript{a}}  \\
  \\
  a. IEMN, UMR 8520 \\
Univ. Lille, CNRS, Univ. Polytechnique Hauts-de-France\\
59000 Lille, France\\
  \\
  b. Escuela Superior de Ciencias Experimentales y Tecnolog\'{i}a,\\ Universidad Rey Juan Carlos, 28933 M\'{o}stoles, Madrid, Spain\\
  \\
  c. Laboratoire Nanotechnologies \& Nanosyst\`{e}mes (LN2)\\
CNRS, Universit\'{e} de Sherbrooke,
J1X0A5, Sherbrooke, Canada\\
  \\
  \texttt{sebastien.pecqueur@iemn.fr} \\
}

\begin{document}
\maketitle

\begin{abstract}
Conventional electronics is founded on a paradigm where shaping perfect electrical elements is done at the fabrication plant, so as to make devices and systems identical, "eternally immutable". In nature, morphogenic evolutions are observed in most living organisms and exploit topological plasticity as a low-resource mechanism for \textit{in operando} manufacturing and computation. Often fractal, the resulting topologies feature inherent disorder: a property which is never exploited in conventional electronics manufacturing, while necessary for data generation and security in software. In this study, we present how such properties can be exploited to implement long-term and evolvable synaptic plasticity in an electronic hardware. The rich topology of conducting polymer dendrites (CPDs) is exploited to program the non-ideality of their electrochemical capacitances containing constant-phase-elements. Their evolution through structural changes alters the characteristic time constants for them to charge and discharge with the applied voltage stimuli. Under a train of voltage spikes, the evolvable current relaxation of the electrochemical systems promotes short-term plasticity with timescales ranging from milliseconds to seconds. This large window depends on the temporality of the voltage pulses used for reading, but also on the structure of a pair of CPDs on two electrodes, grown by voltage pulses. This study demonstrates how relevant physically transient and non-ideal electrochemical components can be exploited for unconventional electronics, with the aim to mimic a universal property of living organisms which could barely be replicated in a silicon monocrystal.
\end{abstract}

\raggedright
\keywords{conducting polymer dendrites \and fractional-order \and memristive capacitors \and evolvable electronics}

\newpage 

\justifying

\section{Introduction}

\begin{figure}[!h]
  \centering
  \includegraphics[width=1\columnwidth]{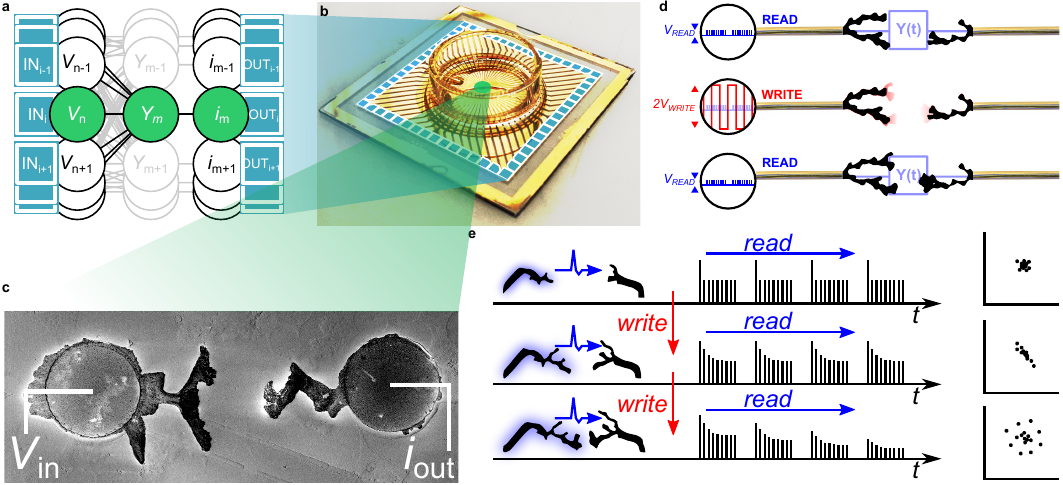}
  \caption{\textbf{Motivations for Evolutionary Neuromorphic Hardware $\vert$ a,} Schematics of a layer of interconnected electrochemical devices where the admittances between input and output contacts in an array condition the transport of a signal, sent by transient voltages on some inputs, and projected as a transient current on some outputs. \textbf{b,} Photograph of a micro-electrode array (from Janzakova \textit{et al.}) \cite{Janzakova2021a} which could be exploited as a hardware playground for evolutionary computation.\cite{Scholaert2024}. \textbf{c,} Scanning Electron Microscope picture (inverted colors) of conducting polymer dendrites evolving between two electrodes on a micro-electrode array (electrode diameter: 30~$\upmu$m). \textbf{d,} Ability for an evolutionary CPD to iterate read-{\&}-write cycles exclusively by voltage activation: to be read at it juvenile state with low amplitude transient voltages, to be written with higher amplitude transient voltages by promoting morphogenesis, to be read at a more mature state with low amplitude transient voltages and so on. \textbf{e,} Objective of this study to demonstrate that CPD morphogenesis allows higher information projection by maturing upon growth, promoting wider windows of synaptic plasticity conditioned by their electrochemical impedance, characteristic of multiple shapes of electrogenerated conductive dendrites.}
  \label{fig:fig1}
  \end{figure}

The current context of electronics manufacturing calls for a revision of the definition of its elementary building blocks. Reasons for these revisions rely on both (1) performances of computing architectures to process information in an unconventional way,\cite{Mehonic2022} and (2) the environmental footprint of electronic hardware micro-fabrication which keeps increasing with anthropological activities.\cite{Wang2023} For both their environmental footprint and their computing performances,\cite{He2019, Teo2023,Guo2013,Jadoun2021,Tropp2021,Kenry2018} organic semiconductors are a key material to implement natural concepts reminiscent of living organisms: one of them is structural morphogenesis. Morphogenesis is a physical ability of sessile organisms to change their structure over the course of their life by scavenging resources on a local scale and evolving according to their needs at a given time in a given place\cite{Coudert2019,Zadnikova2015}. This resource consumption is sparser and activated by low formation enthalpy. It is also highly sensitive to the availability of local sources of nutrients in an environment.\cite{KarlRitz2004,Boddy2008} Currently, conventional electronics do not aim at using physical transience of hardware architectures: Devices do not heal, fabricate by themselves nor recycle easily.\cite{Robinson2009} As a natural expression of intelligence, morphogenesis in living species is also an information processing mechanism.\cite{Tero2010,Rorot2022} It combines the stochasticity of random walks with the determinism of the physical processes triggering their growth and helping living systems converge in their lifetime. It is therefore a computational resource which promotes a classifier's ability to infer environmental patterns in real time and recursively identify threats and opportunities for their development. Applied to electronic systems, hardware structural adaptability can therefore greatly reduce power consumption allocated to computing and information storage with the support of structural adaptation.\cite{Stefatos2006,Lopez2014,Janzakova2023}
It is well accepted that living organisms' intelligence does not entirely rely on topological plasticity, and that a learning system also relies on the mobility of information messengers to activate specific functions in a recurrent nodal system.\cite{Parisi2018} Particularly so for the brain, synaptic plasticity between physically-connected neurons dictates the ability to control information transport and storage on a local scale.\cite{Citri2007,Korte2016} Synapses use mechanisms that are inexistent in silicon: they rule the traffic of multiple information carriers composed of hundreds of different neuromodulators in each elementary synaptic junction.\cite{Dermietzel1993,Roerig2000,Hyman2005} Meanwhile, complementary metal-oxide semiconductor (CMOS) transistors only conduct electrons on two energy bands.\cite{Sze2006} Because each neuromodulator has a non negligible mass and specific chemical reactivity as an ion or as a molecule, its conduction is dictated by diffusion and chemical affinity.\cite{BorrotoEscuela2024,Nicholson2001,Rice1985} Yet, electron states in silicon are governed exclusively by electrostatics.\cite{Sze2006a} The requirements of analog circuits are drastic to master their physical implementation in an electronic circuit: voltage-linear and non-linear devices have specific geometries that must be considered to assess their effects on the local information transport.
\cite{Weste1985,IRDS2023MoreMoore}\\ 
Among passive voltage-linear elements, ohmic resistors and capacitors are the main components of analog electronics. Inductors can also be used to tune the reactance of a circuit, but are much more difficult than capacitors to co-integrate in a small-scale circuit as a thin-film technology.\cite{Sullivan2016,Ondica2023} Ceramic capacitors in an analog circuit obey a classical Debye dielectric relaxation.\cite{Sakabe1997,Coffey2004} If synapses have often been compared to capacitors and are a perspective to be implemented as a neuromorphic technology through memcapacitors,\cite{Mannan2021,Yang2019,Milo2020} the relaxation of biological junctions is far from the behavior of a dielectric relaxation: Experimental characterization of intercellular impedance has evidenced their non-ideality, as they were shown to embed an electrochemical impedance which does not exist as a building block of finite and conventional electronic elements,\cite{Ranck1963,Bedard2013,Bedard2022} but is reminiscent of many phenomena characteristic of interfacial electrochemical relaxations.\cite{Gateman2022,CuervoReyes2015,Bisquert2001} Among them, constant phase elements (CPEs) are intrinsic to diffusion-limited electrochemical relaxations:\cite{RezaeiNiya2016,Yoon2014,Kant2008} the value of their parameters is related to the porosity of a 3D electrode,\cite{Jorcin2006,Jurczakowski2004,Rammelt_1990} or a heterogeneous distribution of the surface potential on a surface area interfacing an electrolyte.\cite{Jorcin2006,Martin2011} 
Not exploited in conventional electronics, such non-ideality in a diffusion-driven relaxation can be a source of enrichment in a neuromorphic architecture. Fractional-order memristive devices exploit their inherent anomalous decay of current or voltage to observe particularly long volatile states.\cite{Petras2009,Venkatesh2023,Li2023,Yan2023,Wu2023,Fortulan2024,Li2024} Tuning the fading-memory time window of synapses is a key element of the fractional-order leaky fire-and-integrate neuron model.\cite{Teka2014,Teka2017,Teka2018} The fractional character in the non-ideal relaxation is controlled by the dispersion coefficient $\upalpha$ characteristic of the non-ideality of a CPE impedance, function of an input signal frequency $\upomega$ as Z(j$\upomega$)~=~$\frac{1}{\textit{Q}(j\upomega)\textsuperscript{$\upalpha$}}$, \textit{Q} being the anomalous capacitance of an electrochemical interface and j\textsuperscript{2}~=~-1.\cite{H.Balaguera2024} The ability to change independently $\upalpha$ from \textit{Q} in an electrochemical junction offers two degrees of freedom to control synaptic plasticity in a memristive capacitor.\\
Recent impedance spectroscopy studies on conducting polymer dendrites (CPDs) have revealed the non-ideality of their relaxation to be strongly correlated to their morphology.\cite{Baron2024,Baron2024a} Multiple relaxations rule their transient behavior in an electrical signal transport, and as CPDs can undergo structural evolution upon voltage activation,\cite{Janzakova2021a,Janzakova2021b} so can their circuit elements, including their pseudo-capacitance \textit{Q}$_{i}$, independently from their dispersion coefficient $\upalpha_{i}$. This study investigates on the possibility of CPDs to exploit structural changes as a programing resource in a neuromorphic device (Fig.\ref{fig:fig1}) from an advanced transient study of the complex current dynamics emulating short- and long-term plasticity in response to voltage stimuli.
  
\section{Experimental Section}

\subsection{Materials and Methods}

The growth conditions, including the involved chemical species, the free standing gold wires and the plastic container are identical to the conditions used in the previous impedance studies:\cite{Baron2024,Baron2024a} 10~mM of 1,4-benzoquinone (BQ) were sequentially introduced with 10~mM of 3,4-ethylenedioxythiophene (EDOT) monomers in a 1~mM concentrated sodium polystyrene sulfonate (NaPSS) electrolyte. The growth voltage is applied using a B1500A Semiconductor Device Analyzer instead of a Solartron Analytical (Ametek) impedance analyzer. The Electrochemical Impedance Spectroscopy (EIS) data shown in Fig.\ref{fig:fig2}.c was obtained using the Solartron impedance analyzer with an AC magnitude of 50~mV\textsubscript{rms} and a 0~V\textsubscript{DC} bias.

\subsection{Transient Current Measurements}

The current measurements shown throughout this study were obtained using a B1500A Semiconductor Device Analyzer and its Waveform Generator Fast Measurement Units (WGFMUs). The WGFMU channels were programmed using the WGFMU instrument library and routines developed for this purpose. The routines mainly involve the writing of sequences of voltage pulses in the first channel while the second channel is set to 0~V. The sequence can contain a single pulse, a continuous train of pulses, or a series of trains of pulses interrupted by periods of rest, all with controllable duty cycle and frequency. Unless specified otherwise, the signal is unipolar and has an amplitude of 50~mV. The specified rising time is 10~ns, and the sampling rate is defined as 100~kHz for an averaging time of 10~$\upmu$s - 10~ns. For all the measurements, the current range was chosen as 10~$\upmu$A. After the data is extracted from the WGFMU, a post-processing Python script is used to remove the first point at each transition time in the current, as it contains a high-frequency non-electrochemical relaxation which affects the envelope of the signal (this non-electrochemical relaxation is either due to the dielectric relaxation of the electrolyte caused by the electrode geometric capacitance or by instrumental limitations including the contribution of the coaxial cable in the operating environment). The envelope itself is obtained by extracting the maximum value of each period. All current measurements are performed on CPDs after they were polarized to 0~V for at least eight seconds in order to ensure similar initial conditions.

\subsection{Current Modelling}

The parameter values of the equivalent circuit shown in Fig.\ref{fig:fig3}.e have been determined by fitting the experimental decay of the currents to the mathematical equation $ i(t) = \frac{VG}{RG+1}(1+RGE_\upalpha[-(\frac{t}{\uptau})^\upalpha])$\cite{Hernandez_Balaguera_2020} using a MATLAB routine\cite{Podlubny2012} for evaluating the intrinsic relaxation pattern in time-domain of the CPE; i.e., the Mittag-Leffler function $E_\upalpha[-(\frac{t}{\uptau})^\upalpha]$. Specifically, the following procedure is performed: (i) $\upalpha$ and $\uptau$ are directly determined by fitting the transient-voltage current response $i(t)$; (ii) $G$ is calculated from the value of $i(0^+)$ ; (iii) $R$ can be found now from $i (\infty) \simeq \frac{VG}{RG+1}$; and finally, (iv) $Q$ is obtained by using the theoretical expression of the characteristic time constant, $\uptau = (\frac{RQ}{RG+1})^{\frac{1}{\upalpha}}$.

\section{Results}

\subsection{Evidence of a non-ideal capacitive charge/discharge in CPDs}

\begin{figure}
  \centering
  \includegraphics[width=1\columnwidth]{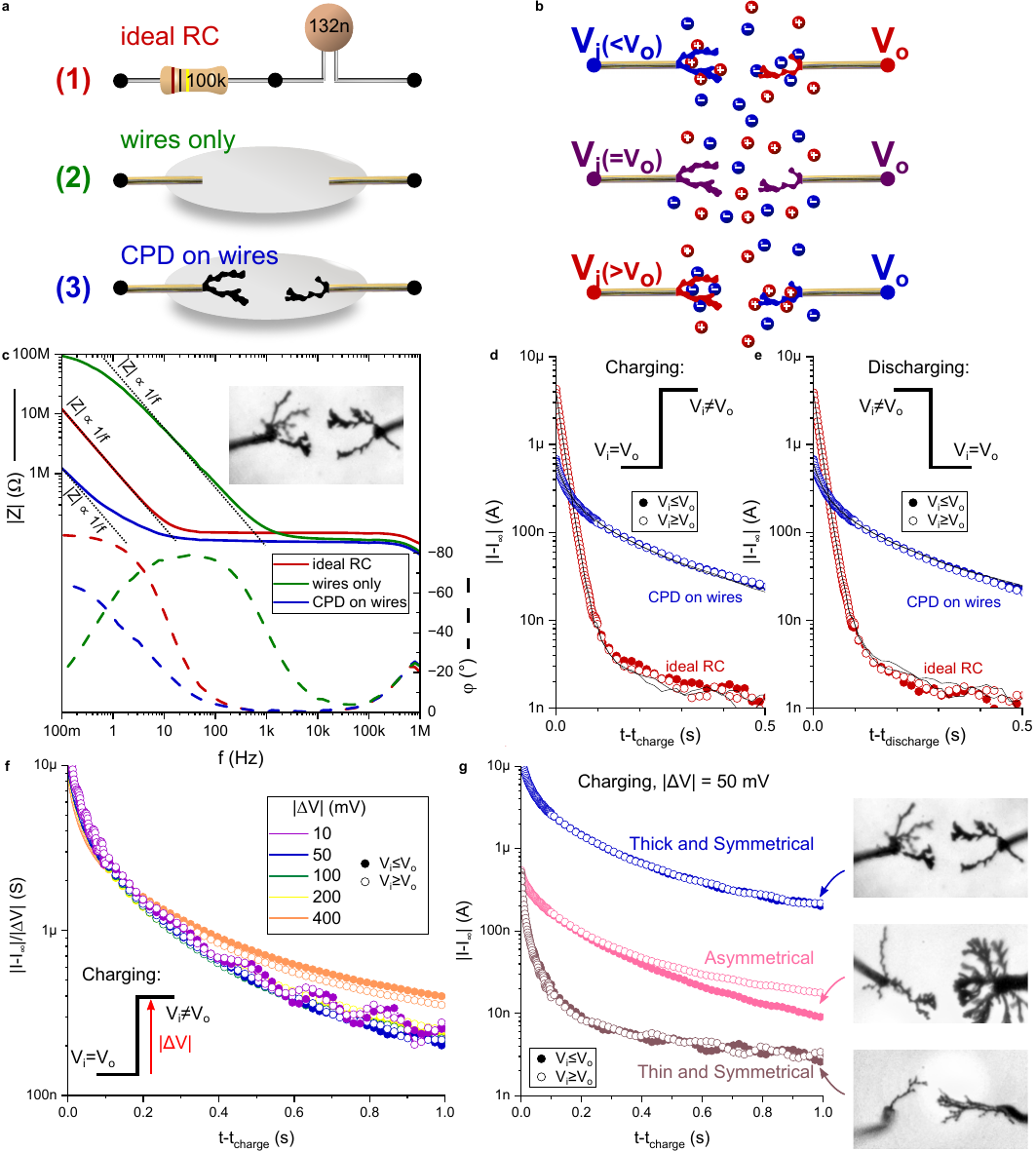}
  \caption{\textbf{Non Ideal Character of the Dynamical Current of a CPD under an Applied Voltage Step $\vert$ a,} Three different experimental setups: a 100~k$\Upomega$ resistor in series with a 132~nF ceramic capacitor (1), naked gold wires immersed in an electroactive aqueous electrolyte before CPD morphogenesis (2) and after morphogenesis (3). \textbf{b,} Schematic representation of the ion charge carrier accumulation on CPDs' morphologies according to an applied DC voltage bias. \textbf{c,} Bode diagram for the impedance of the three different setups depicted in Fig.\ref{fig:fig2}.a, with a microscope image of the characterized CPDs' topology as inset. \textbf{d-e,} Current versus time measurements of the three different setups characterized in Fig.\ref{fig:fig2}.c, for either a charge (\textbf{d}) or a discharge (\textbf{e}). \textbf{f,} Voltage non-linearity window for the dynamic current output from the CPDs characterized in Fig.\ref{fig:fig2}.c, upon a charge. \textbf{g,} Current versus time measurement for the positive and negative charge of CPDs with different morphologies.}
  \label{fig:fig2}
  \end{figure}

A first investigation was performed to quantify the effect of CPDs' non ideality in the time domain, formerly characterized in the frequency domain by EIS.\cite{Baron2024,Baron2024a} To do this, three experimental setups are considered in this study: the free-standing CPD setup, the same setup without CPD, and a conventional serial RC circuit displaying a characteristic time constant that is comparable with the apparent dynamics of the CPD system (Fig.\ref{fig:fig2}.a).
First, while the former two are electrochemical capacitors, the latter uses a dielectric like a ceramic capacitor. In both electrochemical systems, the Helmholtz layer structuring the equivalent capacitance of the interconnect may vary substantially depending on the different ions involved in the electrochemical process, and may depend on the nature of the interfaces they charge on. Specifically, NaPSS has small alkali cations and larger polyanions. Depending on the difference of affinity that each ion may have with the charging material (cations may be transported through the OMIEC but not necessarily PSS\textsuperscript{-}), different charging behaviors can be expected, at the opposite of a Debye capacitor where only electrostatic forces are involved. Second, the structure of the CPD electrodes is morphologically very complex. The frontier defining the Helmholtz layer at the interface between electrolyte and electrode is expected to be very heterogeneous at different scales in the material. Particularly, two CPDs with comparable interface areas may behave differently depending on such heterogeneity, supposedly conditioned by the dendricity of the CPD electrodes at various scales and for different ions. Scholaert and coworkers evidenced that thanks to such asymmetry, the ionic signals of a transient wave depends strongly on the polarity of the wires holding a CPD.\cite{Scholaert2022} We therefore consider two different modes for both the discharge and the charge under a given abrupt change of an applied voltage bias between both CPDs (Fig.\ref{fig:fig2}.b). Here the terms charge and discharge are not defined according to the ionic profile evolution on the electrodes, but by the polarity of these electrodes which promotes ion accumulation relative to their resting potential state. Therefore, a charge designates the electrostatic state of a CPD when a potential difference is applied between the two gold wires, while a discharge describes the state when such potential difference is removed (Fig.\ref{fig:fig2}.b).\\
When EIS is performed, all three systems behave in a similar way as shown in Fig.\ref{fig:fig2}.c. The 100~k$\Upomega$ ohmic resistor in series with four parallel 33~nF ceramic capacitors displays a cutoff frequency of 20~Hz and the impedance modulus features a one decade-per-decade decay with frequency, over two decades of frequency. A 2~Hz cutoff frequency for the CPDs can be extrapolated by the x-coordinate of the intercept of the |Z|$\propto$1/\textit{f} tangent and the high-frequency plateau. However, the slope of log|Z| with log(\textit{f}) never strictly reaches -1 for frequencies above 100~mHz, which is a feature often seen with diffusion-limited capacitors.\cite{Pruneanu2008,Chulkin2017} This feature is attributed to the presence of the CPDs and not to the experimental setup, as the same wires in the same electrolyte environment but without CPD (prior to growing them in the electrolyte) exhibit a substantially different spectrum with a 2~kHz cutoff frequency. The deviation from the |Z|$\propto$1/\textit{f} ideal case appears noticeably more pronounced for the naked wires than for the RC circuit, but far less than after growing dendrites. This is a first evidence that CPDs are dominated by non-ideal relaxations. In the time domain (Fig.\ref{fig:fig2}.d,e), the typical response of a CPD to a charge and a discharge under a bias voltage or the opposite bias is measured. In the case of the RC circuit or of a CPD for which structural asymmetry was not particularly conditioned by the growth, charges and discharges do not seem to depend on voltage polarity (overlapping of the open and filled scatters of the data series displayed in Fig.\ref{fig:fig2}.d,e). Regardless of the voltage polarity, both cationic and anionic charge carriers drift to and from different electrodes, respectively during charging and discharging: the sign of the applied voltage promotes no selectivity in this regard (Fig.\ref{fig:fig2}.b). However, charge carriers having different dynamics would result in polarity-dependent current decays if electrodes were chemically or structurally very different. It is important to stress that despite CPDs being structurally different (inset picture in Fig.\ref{fig:fig2}.c), it does not seem to lead to an obvious difference in decay that would depend on the voltage polarity.
Also in both cases for the RC circuit and the CPDs, comparable dynamics are observed for the charge and the discharge between |$\Updelta$\textit{V}|~=~10~mV and no voltage bias. The goal of this comparison is to verify whether the processes of ion penetration in and escape from CPDs exhibit any hysteresis, which would indicate a greater hindrance to ion migration in one process compared to the other. No such difference is observed in Fig.\ref{fig:fig2}.d,e and confirms that under such stress, the apparent pseudo-capacitance limits the CPD charge and discharge within this range of voltage. The behaviors of the RC circuit and CPDs, however, are quite distinct. While Fig.\ref{fig:fig2}.c shows that their cutoff frequencies are comparable, their dynamics differ in two important ways. First, the current flowing through the CPDs continues to decay after the current of the ideal RC circuit, despite the ideal RC circuit having a higher current than the CPDs at the start of the voltage change (comparison between the red and blue scatters in Fig.\ref{fig:fig2}.d,e). Second, log|\textit{I} - \textit{I}\textsubscript{$\infty$}| does not show any linear relationship with \textit{t} - \textit{t}\textsubscript{charge} or \textit{t} - \textit{t}\textsubscript{discharge} for the CPDs, unlike the ideal RC circuit, until reaching the noise level (Fig.\ref{fig:fig2}.d,e). This feature in particular is a clear indicator that the CPDs do not follow an exponential decay under a voltage step like a conventional capacitor.\\
Different amplitudes of voltage steps have been tested in order to estimate the small signal window of CPDs. Fig.\ref{fig:fig2}.f shows that at 200~mV and below, the resistance decay is comparable. At 400~mV, however, the charging dynamic differs from the others, possibly because of a faradaic contribution in the output current. Meanwhile, at small amplitudes such as 10~mV, oscillations become visible. Therefore, the |$\Updelta$\textit{V}| range of 50–200 mV is acceptable to operate a CPD in its voltage-linear region without limiting the signal to processes other than the impedance involved in the ionic charge/discharge on the CPDs. To quantify the effect of CPD morphologies on a current transient within these boundaries, three different growths have been characterized: two symmetric and thick dendrites made in standard conditions (100~Hz, 50\% duty cycle), two thin dendrites (300~Hz, 50\% duty cycle) and two asymmetrical dendrites (100~Hz, 40\% duty cycle). For each, the current responses to two voltage steps, one of 50~mV, the other of {\textminus}50~mV, were measured. For symmetrical dendrites, charges on both sides have nearly identical dynamics. The thin dendrites, with reduced surface area, show a lower current with a short time constant. On the other hand, the thicker dendrites show a high current with a slower dynamic. As for the asymmetrical system, a clear difference exists between the two charges. Contrary to the other morphologies, in this case, the thinner side gives a higher charging current than the bulkier side. Following the study on CPDs' impedance, it is now clear that symmetrical CPDs' impedance measurements are valid in this range of voltages to study these objects within their voltage linearity range. Remaining within these boundaries allows us to use voltage-invariant elements such as constant-phase elements responsible for fractional-order capacitance in the time-domain characteristics.

\subsection{Behavior under unipolar and periodic voltage pulses}

\begin{figure}
  \centering
  \includegraphics[width=\columnwidth]{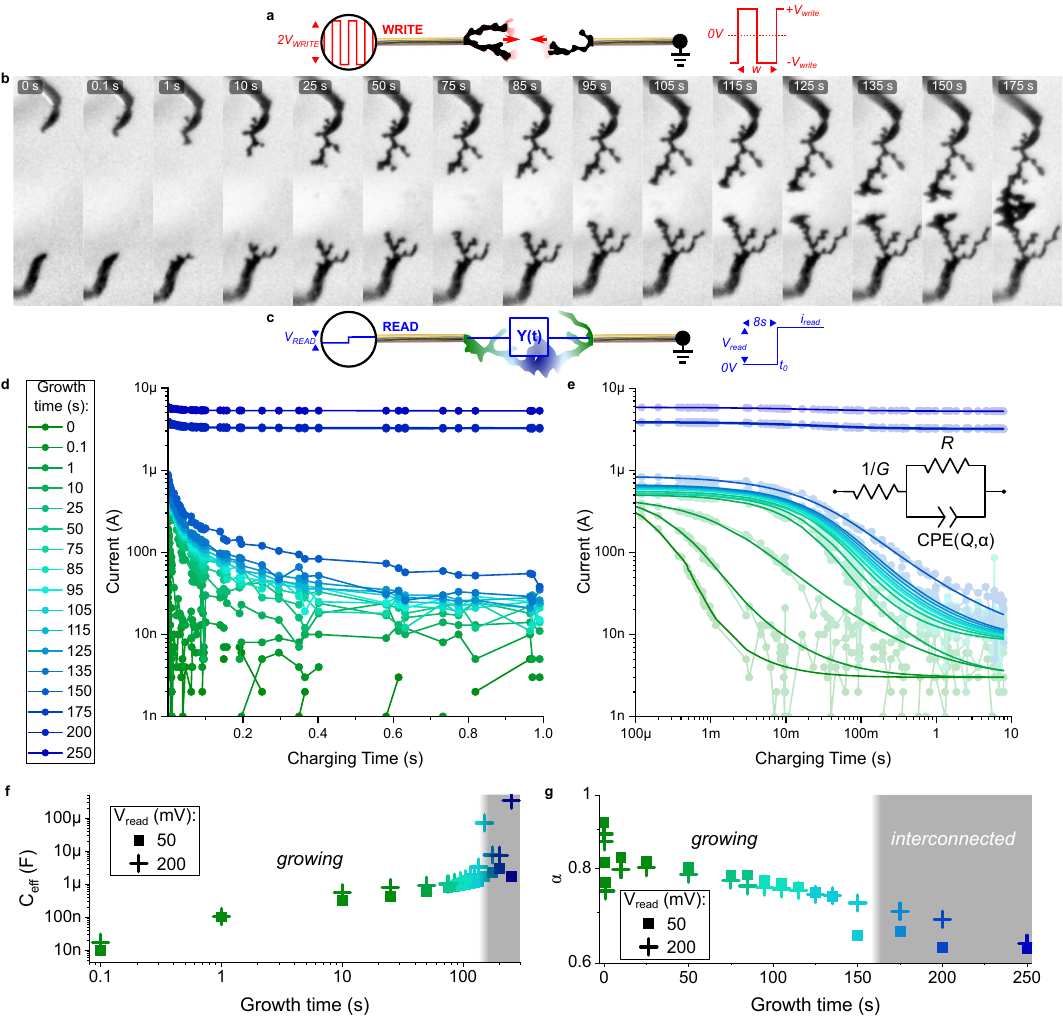}
  \caption{\textbf{Evolution of CPDs' Circuit Paramenters upon Morphogenesis $\vert$ a,} Setup for writing on a CPD topology with voltage pulses. \textbf{b,} Photographs of a CPD at multiple evolution stages. \textbf{c,} Setup for reading a CPD current under a voltage step. \textbf{d-e,} Transient current of the CPD growth in Fig.\ref{fig:fig3}.b in logarithmic (\textbf{d}) and linear (\textbf{e}) scale of time from the start of the step. \textbf{e-g,} Fittings the current response of a CPD with the Mittag-Leffler function as a time-domain expression of the CPE containing Voigt equivalent circuit: Fits are printed as solid lines in Fig.\ref{fig:fig3}.d, over experimental data scatters. \textbf{f-g,} Evolution of the fitted parameters \textit{Q} and $\upalpha$ along with growth.}
  \label{fig:fig3}
  \end{figure}

Previous studies in the frequency domain have shown that the impedance spectrum of the CPD-on-wire system evolves significantly at the beginning of the growth\cite{Baron2024,Baron2024a}. Specifically, at least three relaxations were apparent and were shown to change during the growth. A high frequency relaxation ($\uptau$\textsubscript{3}~$\simeq$~1~$\upmu$s), an intermediate frequency relaxation ($\uptau$\textsubscript{2}~$\simeq$~10~ms) and a low frequency, highly resistive relaxation ($\uptau$\textsubscript{1}~$\simeq$~10~s). The time constant of the second relaxation, likely corresponding to the charge of the electric double layer of the CPD, gradually becomes longer as more material is deposited. This particular relaxation was shown to vary in both \textit{Q}\textsubscript{2} and $\upalpha$\textsubscript{2}, especially when comparing to different morphologies (thin and bulky CPDs), while $\upalpha$\textsubscript{1} mostly remained close to ideality. However, contrary to EIS data, measurements in the time domain under unipolar voltage stresses allow studying the relaxations strictly under two regimes where ions are either accumulated on CPDs with a bias or diffuse back in the absence of voltage bias. Being able to modify the equivalent circuit parameters of a single dendrite in real time using low voltage gives access to a certain range of non ideal dynamics that needs to be evaluated. To ensure that CPDs can store electrically accessible information in their morphology, their transient current is studied at different degrees of maturation. Voltage steps of 200~mV lasting eight seconds were applied to the input CPD in order to follow the evolution of its transient response, as it goes through the different stages shown in Fig.\ref{fig:fig3}.b. The first frame shows two uncoated wires, while the following frames show the growth until completion. Each frame corresponds to a measurement. The time specified on each frame is the electropolymerization time, which took around 175~s to complete. Additional measurements were performed after completion. As 200~mV is below the voltage necessary to oxidize EDOT, CPDs are not modified during the measurements while their state is being read.
The current measurements for each stage are compared in Fig.\ref{fig:fig3}.d. The progression is consistent with the results of the impedance study: the time constant appears to increase during growth, while the level of current also increases because of the increasing surface area of the CPDs. A change in morphology directly translates to a change in charge dynamics, although most of the change occurs in the first ten seconds. The semi-log graph does not show a linear dynamic, a characteristic of an ideal capacitor, which is a proof of the non ideal character of the CPD charge. In Fig.\ref{fig:fig3}.e particularly, the log-log graph allows to see the considerable impact the very first oligomers of PEDOT:PSS have on the dynamic of the current charge after only 100 ms of electropolymerization. This confirms the results obtained through impedance spectroscopy by Baron et al\cite{Baron2024a}. Here, during the first ten seconds of the charge, only the second slowest relaxation is noticeable. Because of the high impedance at low frequencies, very little current flows after the first second, and the remaining signal becomes noisy. Despite the noise, a slight drift can be observed happening over the last few seconds, suggesting the presence of the slowest timescale relaxation that was evidenced by the distribution of relaxation times\cite{Baron2024a}. The apparent conflict between both techniques' results is likely to be attributed to the limitations of both instruments, which have fundamentally different stimulation protocols, integration limits and sampling methods. In the present study, this highly resistive relaxation of a long timescale will be neglected to avoid risks of overfitting the current responses. Additionally, the high frequency relaxation observed at the microsecond scale is not visible in the transient response, as its timescale is so short that it only really exists in the first sampling point. Considering the sampling time is 10~$\upmu$s, the microsecond-long relaxation has already reached its final value by the time the second sampling point is integrated. It is likely that this specific relaxation, initially thought to be the dipolar relaxation of the solvent, is in fact caused by the capacitance of the cables. Even after the CPDs are connected, a single relaxation is still clearly visible, which is not unexpected as side branches of CPDs were shown to continue growing after completion, a sign that ions still migrate to the surface/within the material. An inset to Fig.\ref{fig:fig3}.e shows the equivalent circuit chosen to model the system and extract the electrical parameters of the CPDs, consisting of a single constant phase element.
The dispersion coefficient $\upalpha$ and the effective capacitance \textit{C}\textsubscript{eff} were extracted from the transient response at each stage of the growth and are displayed in Fig.\ref{fig:fig3}.f-g.\cite{Hirschorn2010} The effective capacitance is a function of the CPE parameters (\textit{Q} and $\upalpha$) and the resistances (\textit{R} and 1/\textit{G}) as defined in a previous study\cite{Baron2024}. It is confirmed that both quantities \textit{Q} and $\upalpha$ are changing during the growth, although they evolve at different timescales. The evolution of the effective capacitance is monotonous, and mostly changes during the first milliseconds of the growth, whereas the dispersion coefficient abruptly changes at the beginning and linearly decreases over time. This evidences that both parameters depend on different physical properties. It is likely that the relaxation modeled in the present study relates to the second relaxation of parameters $\upalpha$\textsubscript{2} and \textit{C}\textsubscript{eff,2} in the impedance study of CPDs under the same configuration\cite{Baron2024a}.
However, when compared to the $\upalpha$\textsubscript{2} and \textit{C}\textsubscript{eff,2} values extracted from the impedance spectra, slightly different dynamics are observed between the two studies. In the frequency domain, two relaxations were modeled, and in the case of the time domain measurements, only the charge of a single dendrite is studied. Moreover, while the same growth parameters were used, they will inevitably yield different morphologies because of the intrinsic stochasticity of CPD growth, which will alter the electrical properties of the dendrites. All these elements could explain the differences observed in both fittings. The two studies however agree on the range of values for $\upalpha_2$, which is centered around 0.8 when the CPDs are fully grown.
Two different amplitudes for the voltage steps were studied (200~mV and 50~mV) to assess whether non linear effects could exist in this range of voltages. Both datasets match and give the same fitting values for $Q$ and $\upalpha$. The values of $Q$ and $\upalpha$ are slightly different for time samples after CPDs are completed when comparing the two voltage stimulations. This is likely to be attributed to the fact that the electrochemical model is inadequate and used beyond its boundaries (as electronic conduction through the dendrites dominates the electrochemical current flows). The key point is that, for uncompleted dendrites, most of the information about the CPD maturity is contained within the first milliseconds of the transient response and that such information is two dimensional, materialized by two circuit parameters: the anomalous capacitance $Q$ and the dispersion coefficient $\upalpha$ of CPDs at a given maturity.\\

\begin{figure}
  \centering
  \includegraphics[width=\columnwidth]{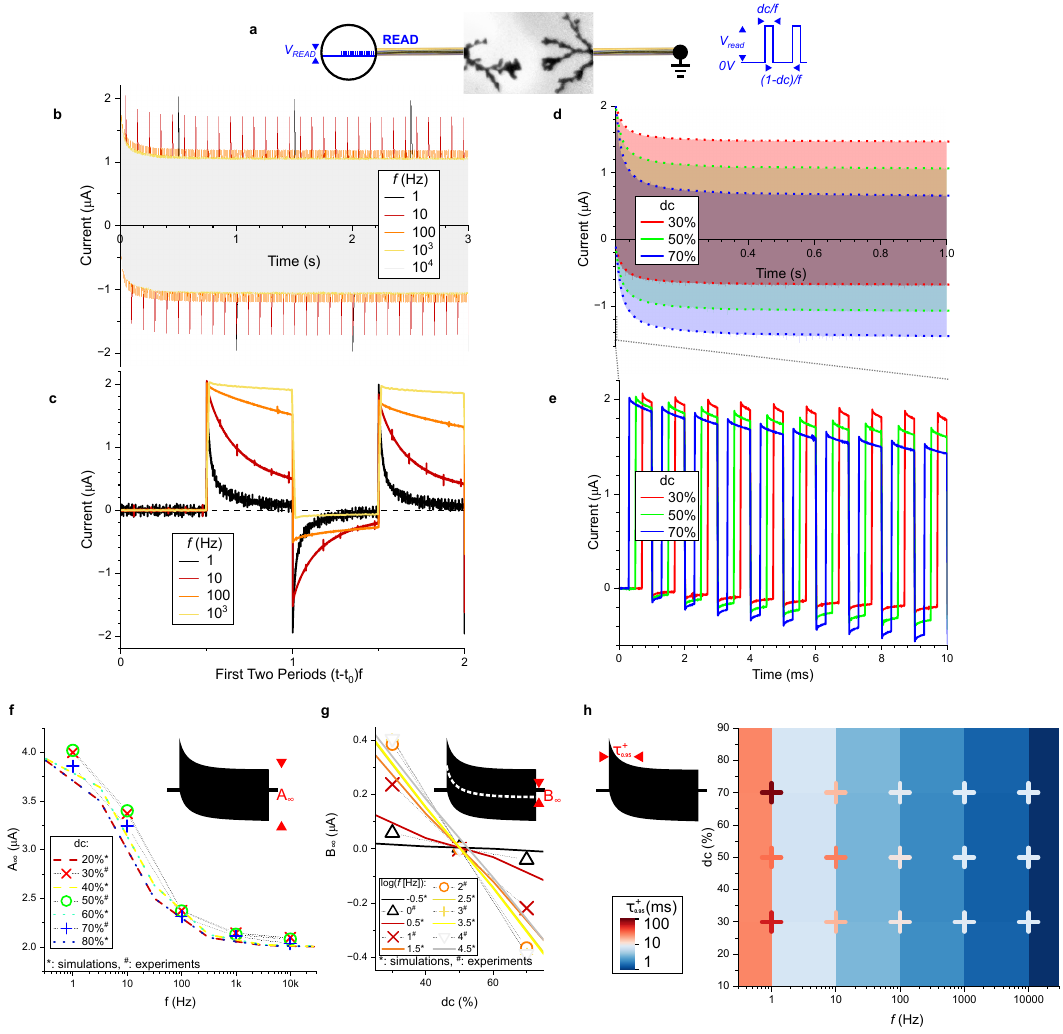}
  \caption{\textbf{Waveform-Parameter Dependencies of CPDs' Current Transient under a Pulse-Voltage Input $\vert$ a,} Setup for reading a CPDs' transient current with an unipolar pulse wave, with a photograph of the CPDs characterized in the following subfigures. \textbf{b-e,} Frequency and duty cycle dependence of the CPDs' current (\textit{f} - \textbf{b,c}) and duty cycle (dc - \textbf{d,e}) of the applied voltage signal. \textbf{f-h,} Signal envelope's dependency with \textit{f} and dc, specifically on the amplitude A of the envelope (\textbf{f}) the position B of the envelope (\textbf{g}) and the apparent time constant at 95\% (\textbf{h}). \textbf{f-h}: Scatters are from experimental data, color map and solid lines are from simulated data obtained from the fitted model.}
  \label{fig:fig4}
  \end{figure}

Because this information fades very quickly, periodically repeating the transience by applying a pulse signal may allow us to visualize over longer timescales the information related to a CPD morphology in a current signal. Therefore, a train of unipolar voltage pulses has been applied and the current response of the CPD system is studied under those conditions (Fig.\ref{fig:fig4}). To investigate the impact of this waveform on the quality of the current signal, different pulse frequencies and duty cycles have been tested for unipolar pulses fixed at a given voltage amplitude of 200~mV. In this study, it is identified that both frequency and duty cycle have an impact on how the information about CPDs' morphology is carried in the signal.\\[3pt]
In this scenario (Fig.\ref{fig:fig4}.a), charges and discharges are occurring periodically, and the information is carried by the envelope of the current, which is sampled every pulse by taking the maximum value over the period. In Fig.\ref{fig:fig4}.b, the effect of the frequency of the pulses is visible in the raw current data. For a frequency of 1~Hz, the envelope is still, as a half second pulse is enough to charge the CPDs almost completely and reach a quasi steady-state. From 10~Hz, a transient can be seen at the beginning of the envelope. The charging of the CPDs is periodically interrupted and ions diffuse away from the electrode during the low level pulses, which slows down the charge, but not enough to completely compensate it. After a few pulses, the CPDs' charge/discharge reaches an equilibrium and the envelopes of the current stay constant. This effect is a neuro-mimetic emulation of short-term plasticity (STP), which is now often described in the literature when capacitive systems are stimulated with fast unipolar pulses of voltage, promoting facilitation on the upper envelope and depression in the lower envelope.\cite{Giordani2017} This equilibrium value for the current decreases as the frequency increases. At the same time, the envelope transient gets narrower and its time constant decreases. However, the initial current value remains the same for all frequencies, as the current is maximum when all charges are free to move in solution with no charges on the electrodes. It should be noted that this type of transient in the envelope is not only observed in CPDs, but is also expected in ideal capacitors. Moreover, it is not necessary for the charge and discharge to have different dynamics for this transient to happen. Because the discharge is interrupted before its completion, some residual mobile charges still remain on the electrode at the start of every charging process, lowering the voltage driving the charge/discharge current and hindering the build up of future charges. While less and less charges can be added to the electrode, more and more charges can be released during the discharge. Eventually, both charge and discharge reach an equilibrium and balance themselves. At this point, the amount of charges leaving the electrode during the discharge is equal to the amount of charges coming to the electrode during the charge. The duration required for this equilibrium to take place depends on the electrode capacitance.The CPDs' high degree of tunability allows to modify when this equilibrium takes place, as well as to change the dynamics of the envelopes.\\[3pt]
The charge/discharge dynamics within the first pulses are visible in Fig.\ref{fig:fig4}.c for the different frequencies. A difference in dynamics between charge and discharge exists initially, even for a 50\% duty cycle train of pulses. The current response was also investigated in terms of duty cycle values, for 30\%, 50\% and 70\% duty cycle. Varying the duty cycle of the signal aims at tuning the charge/discharge balance of both CPDs to promote different ion doping profiles and read CPDs under different conduction states. For a duty cycle of 50\%, the current is balanced and, after the initial transient has disappeared, the low level envelope is symmetrical to the high level one, as shown in Fig.\ref{fig:fig4}.d. For higher duty cycles, the current is shifted negatively. This is because the longer the high level pulse is, the more charges are added to the electrode, lowering even more the resulting current. On the contrary, at a 30\% duty cycle, there are fewer charges on the electrode. Being less screened, the electric field allows more ions to move faster to the surface, resulting in a higher current.\\[3pt]
Because the initial current is the same for the three values of duty cycle, the envelope transient, which is visible in Fig.\ref{fig:fig4}.e, has an amplitude and an apparent time constant which both depend on the value of the duty cycle. Therefore, one has the ability to stimulate the CPDs with signals operating at different frequencies and duty cycles to read their intrinsic morphological properties, contained in the envelope of the current. Metrics such as the magnitude of the current A, its average steady state value B and an estimation of the time constant $\uptau^+_{0.95}$, can also be extracted from the measurements. Fig.\ref{fig:fig4}.f shows the evolution of the magnitude of the current with the frequency, which is maximal for low frequencies, and quickly decreases to the same value at higher frequencies. Meanwhile, the evolution of the average steady state value is presented in Fig.\ref{fig:fig4}.g, and it is shown to evolve linearly with the duty cycle, reaching 0 at 50\% duty cycle. Finally, in Fig.\ref{fig:fig4}.h, the time constant of the transient is studied in terms of both duty cycle and frequency. The higher the duty cycle, and the longer the time constant tends to be. Meanwhile, the higher the frequencies, the shorter the time constant becomes. Those values are compared to the values obtained for the same metrics in simulation. The system was simulated considering a single relaxation. Simulated data of Fig.\ref{fig:fig4}.f-h were obtained from the fitting parameters of a representative transient current pertaining to the CPDs of Fig.\ref{fig:fig3}.b in an intermediate state during the growth (\textit{G}~=~10~$\upmu$S, \textit{R}~=~20~M$\Omega$, $\upalpha$~=~0.75 and $\uptau$~=~25.8~ms). The details of the fitting procedure are provided in the Experimental section. Importantly, we show that the envelope current is stabilized when in a period, the transient current both during the on state and the off event are equal in absolute value, which is significant to obtain and interpret the values of A\textsubscript{$\infty$}, B\textsubscript{$\infty$} and $\uptau^+_{0.95}$. To this end, the use of Mittag-Leffler relaxation pattern $E_\upalpha[-(\frac{t}{\uptau})^\upalpha]$\cite{Mainardi2014}, fractional dynamics approach, and beyond is indeed key to obtain realistic values in accordance with experimental measurements\cite{Metzler2002}. The non-exponential decays are given by $i(\infty) + [i(t^+_0) - i(\infty)]E_\upalpha[-(\frac{t}{\uptau})^\upalpha]$\cite{Hernandez_Balaguera_2020}, where the steady-state current, mathematically expressed as $i(\infty)$, is $\frac{VG}{RG+1}$ and 0~A for the charge and discharge phases, respectively, and $i(t^+_0)$ is the transient term that accounts for the current memory traces intrinsic to the experiment, just after the step changes at the time instant \textit{t}\textsubscript{0}. Note that $i(t^+_0) = i(t^-_0) \pm VG$ (that is, positive or negative sign depending on if one considers the leading or trailing edge of the pulses). $t^-_0$ denotes the time just before the step changes. However, the simulation does not take into account the existence of a transient in the dynamics of charge/discharge. The dynamic of the current decay within a single pulse duration evolves over time, although the overall dynamics of the envelope in a pulse train quickly stabilizes. For a 50\% duty cycle, charge and discharge converge to identical dynamics over time. For a different duty cycle, their dynamics will differ even in the steady state, meaning that different dynamics need to be implemented in the simulation for the high and low level pulses. This behavior, which is not present in conventional dielectric capacitors, is a testament to the peculiar physics governing CPDs.

\subsection{Short- to long-term synaptic plasticity implemented on an evolving CPD}

\begin{figure}[!h]
  \centering
  \includegraphics[width=1\columnwidth]{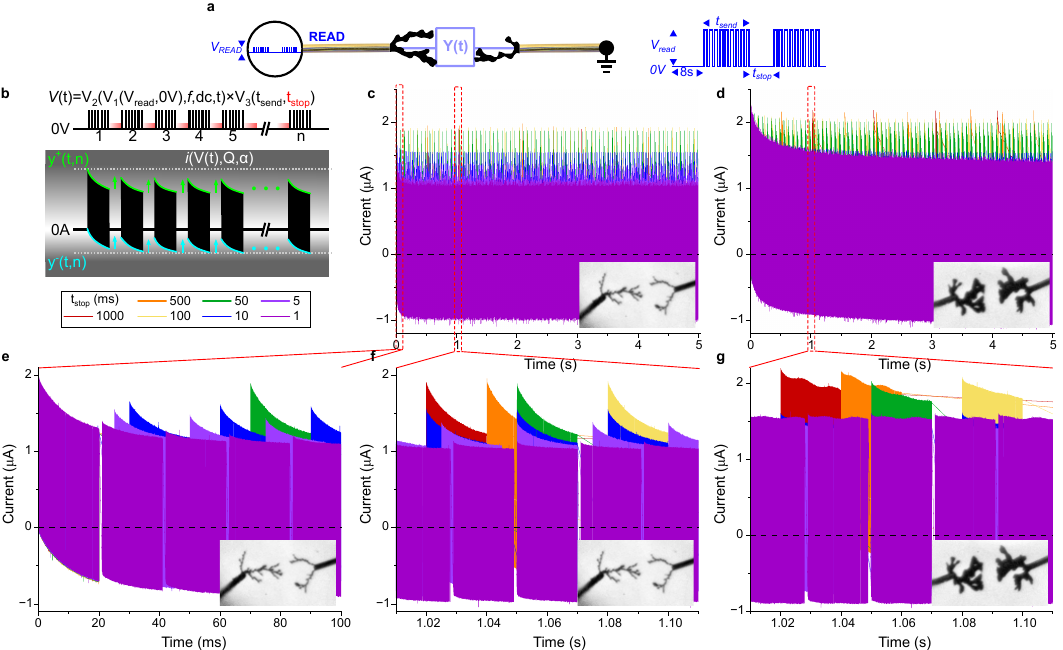}
  \caption{\textbf{Structural Dependency of CPDs' Time Plasticity $\vert$ a,} Setup for reading a CPD transient current with interrupted sequences of unipolar pulse wave. \textbf{b,} Extracting CPDs' envelope signals and information carriers featuring temporal properties comparable to synaptic plasticity. \textbf{c-e,} Current signals, for the different CPDs displayed in their respective inset, for a signal waveform such as \textit{V}\textsubscript{read}~=~200~mV, \textit{f}~=~10~kHz, dc~=~50\%, \textit{t}\textsubscript{send}~=~20~ms for different \textit{t}\textsubscript{stop} (common legend in Fig.\ref{fig:fig5}.b).}
  \label{fig:fig5}
  \end{figure}

Stimulating CPDs with a continuous train of pulses revealed several properties in the output current that are directly related to their electrical parameters. It has previously been shown that these electrical parameters could be changed at will by either growing the CPDs further or changing their shape through the growth parameters. This gives access to a certain range of volatilities by either changing $\upalpha$ with the CPD morphology or $\uptau$ by modifying the unipolar voltage pulse waveform, for which the electrical admittance could be used as a synaptic weight in an analog neuromorphic circuit. Specifically, CPD electrical signals are a function of both their past (the voltage past experience that yielded a specific morphology) and their current environment (which can condition different values of $\uptau^+_{0.95}$). We have seen that for a continuous train of pulses, a transient can appear in the envelope. This transient is of major importance as it contains the dynamical information of the system. However, it quickly fades away when the equilibrium is reached. A new stimulus is shown Fig.\ref{fig:fig5}.a. By introducing periods of rest between the packets of pulses, a local transient is either maintained or the equilibrium is reached and the envelope remains constant.
Based on the results discussed in the previous section, the frequency of the reading pulses was set to 10~kHz to obtain a good resolution for the envelope of the periodic packet of pulses. The duration \textit{t}\textsubscript{stop} presented in Fig.\ref{fig:fig5}.b corresponds to the duration of the rest period. In Fig.\ref{fig:fig5}.c-d, the effect of this duration on the current response is studied for two morphologies: thin (using a growth frequency of 200~Hz) and bulky (30~Hz) CPDs. For both morphologies, a transient of the envelope is only visible for low values of \textit{t}\textsubscript{stop}.  
  Indeed, for high values, the relaxation has enough time to complete before the next packet of pulses. The thin CPDs, when compared to the bulky ones, yield faster dynamics, and therefore quickly lose their transient as $t_{stop}$ is increased. A zoom-in on the first pulses of the thin CPDs is shown in Fig.\ref{fig:fig5}.e. 

  It shows that the initial current value of each train of pulses is increasing with $t_{stop}$. Fig.\ref{fig:fig5}.f-g focus on the dynamics of the current around the first second. At this stage, the envelope transient has mostly disappeared, but the local transient of the packets is still maintained for high values of $t_{stop}$. In fact, two transients are occurring, one is more local and on the timescale of a single packet; the other concerns the whole signal and has a long timescale. The former can be studied using a short term potentiation (STP) rate: for a given packet, this rate can be defined as:
$$r_{STP} = \frac{1}{5}\sum_{i=250}^{255}\frac{y^+_i(T_{packet}) - y^+_i(0)}{y^+_i(0)}$$
The latter corresponds to a long term potentiation (LTP) rate, defined as:
$$r_{LTP} = \frac{1}{5}\sum_{n=T_{packet} - 5}^{T_{packet}}\frac{y^+_{255}(n) - y^+_{0}(n)}{y^+_0(n)}$$

\begin{figure}
  \centering
  \includegraphics[width=1\columnwidth]{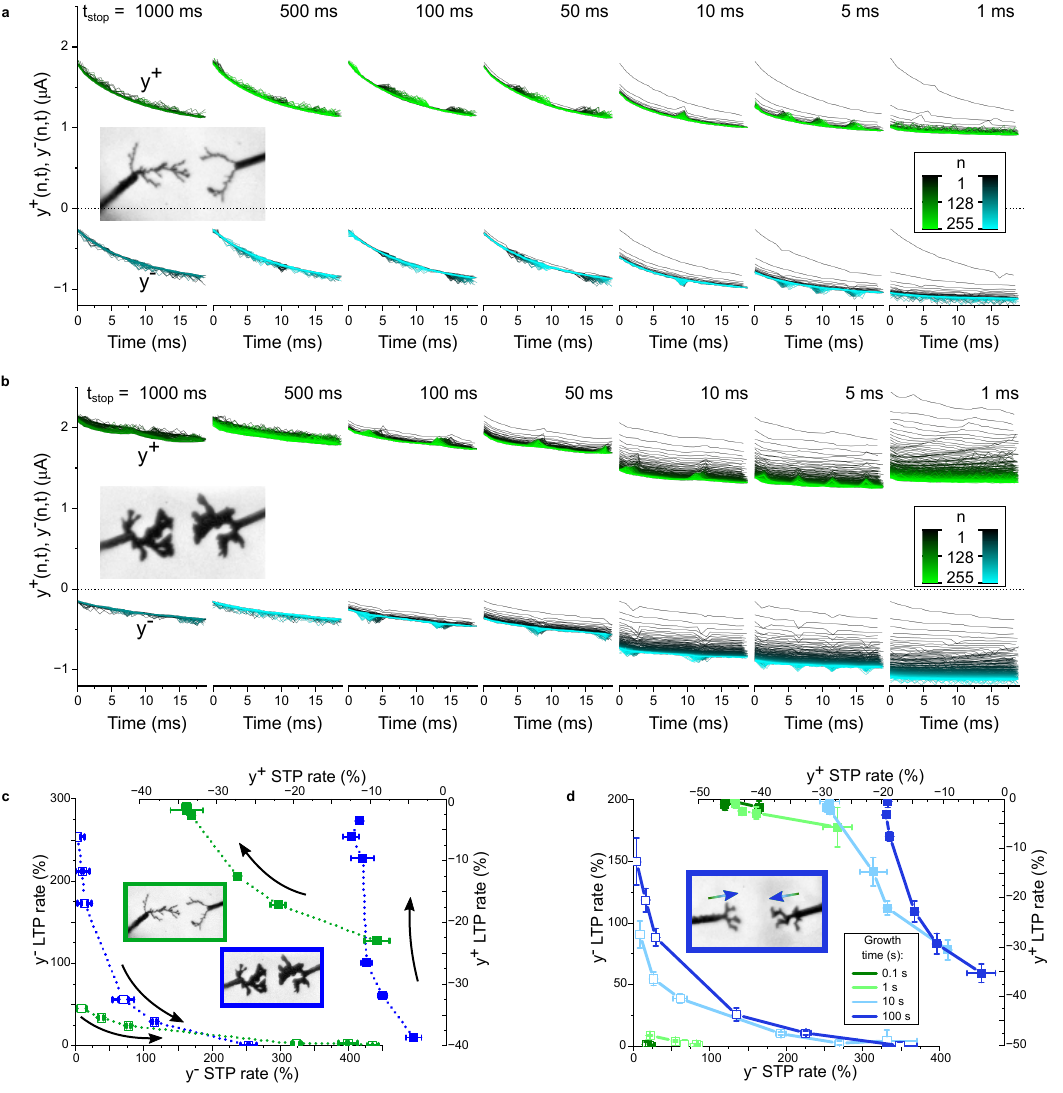}
  \caption{\textbf{Evolution of CPDs' Time Plasticity with Morphogenesis $\vert$ a,b,} Upper (y\textsuperscript{+}) and lower (y\textsuperscript{-}) envelope signals over time from Fig.\ref{fig:fig5}.c,d raw data, for each repeated sequence n in a measurement and for the different CPDs displayed in their respective inset, as a function of t\textsubscript{stop}. \textbf{c,} Long and short term potentiation rates for two different CPD morphologies. \textbf{d,} Long and short term potentiation rates for four different stages of the CPD growth, for a growth frequency of 50~Hz.}
  \label{fig:fig6}
  \end{figure}

Here, $y_i^+$ designates the upper envelope of the i\textsuperscript{th} packet in the current. $T_{packet}$ is the duration of a packet.  
The current responses to the first 255 packets are analyzed, stacked together and compared for each value of $t_{stop}$ in Fig.\ref{fig:fig6}.a-b for both morphologies. The thin CPDs display a steep decrease of current for each packet when compared to the thicker CPDs. Both envelopes (low and high level) show the same behavior. For low values of $t_{stop}$, the transient of the envelope is noticeable at least for the first charge. Afterwards, the system converges to a flatter charging dynamic. The more $t_{stop}$ is increased, the less the envelope's transient is apparent, and the steeper the dynamics become. The transient of the envelope finally disappears around $t_{stop}~=~50~ms$. On the other hand, for the bulky dendrites, this transient is barely visible starting after $t_{stop}~=~100~ms$. The dynamic is also noticeably flatter than in the case of the thinner CPDs. This dataset evidences that the charge/discharge dynamics of the CPDs evolve depending on the resting time $t_{stop}$. Long and short term potentiation rates were extracted for both morphologies and are presented in Fig.\ref{fig:fig6}.c. The data for the thin CPDs are shown in green while the data of the thick CPDs are depicted in blue. The black arrows indicate the direction of increasing $t_{stop}$. For low values of $t_{stop}$, the thinner CPDs
show a high LTP rate. As $t_{stop}$ increases, the LTP rate decreases in favor of the STP rate. The same behavior is achieved with the bulkier CPDs, although the dynamic of achievable STP rates is reduced, while the LTP range is enhanced. By playing with the resting time $t_{stop}$, one mode of potentiation can be used over the other, and the morphology further conditions this type of potentiation. Finally, the potentiation modes are studied for a CPD at different stages of its growth (Fig.\ref{fig:fig6}.d). At distinct electropolymerization times (0.1~s, 1~s, 10~s, 100~s), the potentiation rates are extracted and show that as the CPDs grow, the range of achievable STP rates is reduced, while more control over the LTP is obtained. This happens particularly during the first ten seconds of electropolymerization. A consequence is that for a given value of $t_{stop}$, a signal that did not cause LTP at the early stage of growth can, upon further growth, cause LTP. It is a direct example of an evolvable electronic device that grows, and as a result, shows changing electrical properties which affect signal transport. Such dynamic interconnections could be implemented in a network of nodes and learn in a way similar to the brain. CPDs that fire together, can physically wire together by dynamically growing toward specific nodes in different ways that directly depend on their past history, changing their properties in the process. Two CPDs at different stages of their life can therefore vehicle different electrical signals based on their past exposure to inputs, and once integrated with sensors, have the potential to support, if not replace, conventional software  classifiers.

  \newpage
  
\section{Conclusions}

The study confirms the relevance of using the evolutionary character of CPDs to process information for unconventional electronics. As linear components (below 200~mV), two uncompleted CPDs behave as an imperfect capacitor: their current decays faster in the short term and slower in the long term than with conventional dielectrics used in analog electronics. Beyond their linearity window, they grow under voltage activation (typically 5~V) so their structural change modifies their level of non-ideality in a permanent way. By growing, such evolving devices can tune their filtering properties for spike-voltage transmission on two distinct levels: depending on their morphology, time constants for both short-term and long-term synaptic plasticity can be tuned according to the voltage history engraved in the morphology of a physically-transient topological device. As the process requires only low voltage (5~V or even lower) and low modulations (10~kHz or lower) to write-in and read-out, CPD morphogenesis demonstrates its full potential to be implemented in standardized electronic systems to embed non-conventional information-processing resources, drawing inspiration from biology to use evolving hardware to process and store information. Near-sensor implementation shall be demonstrated in future studies to embed calibration points for non-metrological classification sensing platforms or storing data locally on Internet-of-Things information generators in a non-vulnerable way.

\section*{Acknowledgments}
The authors thank the French National Nanofabrication Network \href{https://www.renatech.org/en/}{RENATECH} for financial support of the IEMN cleanroom. We thank also the IEMN cleanroom staff for their advice and support. This work is funded by ANR-JCJC "Sensation" project (grant number: \href{https://anr.fr/Projet-ANR-22-CE24-0001}{ANR-22-CE24-0001}), the ERC-CoG "IONOS" project (grant number: \href{https://doi.org/10.3030/773228}{773228}) and the Hauts-de-France R{\'e}gion.

\section*{Competing Interests}
The authors declare no competing interests.

\bibliographystyle{natsty-doilk-on-jour}  
\bibliography{ref}  

\end{document}